\documentclass{PoS}%
\usepackage{amsmath}
\usepackage{amsfonts}
\usepackage{amssymb}
\usepackage{graphicx}%
\providecommand{\U}[1]{\protect\rule{.1in}{.1in}}
\ShortTitle{Revisiting the axial anomaly}
\abstract{The axial anomaly is responsible for the masses and mixing of the mesons $\eta$ and $\eta'$. An open question is if (and to what extent)  it affects also other hadrons. We show that anomalous terms can be important to understand the spectroscopy of the pseudotensor mesons $\eta_2 (1645)$ and $\eta_2 (1870)$. In fact, pseudotensor mesons belong to a so-called heterochiral  multiplet, for which a quadratic mixing term between nonstrange and strange isoscalar members arises. On the contrary,
for so-called homochiral  multiplets, such as the ground-state (axial-)vector and tensor mesons, this mixing is not possible, hence one can easily understand why the isoscalar members of these multiplets are almost purely nonstrange and strange, respectively.
Moreover, the axial anomaly can be also coupled to baryons (within the mirror assignment), and thus it helps to explain the large decay width $ N^{\ast }(1535)\rightarrow N\eta$  and to clarify which baryons are chiral partners. }
\FullConference{XVII International Conference on Hadron Spectroscopy and Structure - Hadron2017\\
		25-29 September, 2017\\
		University of Salamanca, Salamanca, Spain}

\title{Revisiting the axial anomaly for light mesons and baryons}
\author{\speaker{Francesco Giacosa}\thanks{A footnote may follow.}\\Institute~of~Physics, Jan~Kochanowski~University, ul.~Swietokrzyska 15, 25-406 Kielce\\Institute~for~Theoretical~Physics, Johann~Wolfgang~Goethe~University,
Max-von-Laue-Str.~1, 60438 Frankfurt~am~Main\\E-mail: fgiacosa@ujk.edu.pl }
\begin{document}
\section{Introduction}

In the chiral limit, the QCD Lagrangian contains a $U(1)_{A}$ symmetry, which
is broken by quantum fluctuations: this is the so-called axial anomaly
\cite{thooft}. Its role is necessary to understand the properties of the
mesons $\eta\equiv\eta(547)$ and $\eta^{\prime}\equiv\eta^{\prime}(958)$
(especially the mass of $\eta^{\prime}$ and the large mixing between
nonstrange and strange contributions) \cite{feldmann}. Yet, what about other
mesons? And what about the role of the axial anomaly for baryons? In these
proceedings, we report on two recent works in which the role of the axial
anomaly was studied in the mesonic and baryonic sectors
\cite{anomaly,baryonanomaly}.

In\ sec. 2, based on Ref. \cite{anomaly}, light meson nonets are grouped into
chiral multiplets and classified according to their chiral transformations:
there are \textquotedblleft heterochiral\textquotedblright\ mesons (such as
the (pseudo)scalar states) and `homochiral' mesons (such as (axial-)vector
states). For all heterochiral multiplets, the axial anomaly allows for mixing
of nonstrange and strange isoscalar members, as well known for the $\eta$ and
$\eta^{\prime}$ mesons mentioned above. In addition, such a mixing seems to be
realized also for heterochiral pseudotensor mesons \cite{adrian}. On the
contrary, this anomalous mixing is not possible for \textquotedblleft
homochiral\textquotedblright\ multiplets: one can easily understand why the
vector mesons $\omega(782)$ and $\phi(1020)$ as well as the tensor mesons
$f_{2}(1270)$ and $f_{2}^{\prime}(1525)$ are (almost purely) nonstrange and
strange, respectively.

In\ Sec. 3, based on Ref. \cite{baryonanomaly}, the axial anomaly is used to
understand the anomalously large decay $N(1535)\rightarrow N\eta$ (much larger
than than what flavor symmetry naively predicts) as well as the decay
$\Lambda(1670)\rightarrow\Lambda(1116)\eta$. Namely, the axial anomaly allows
for a chiral anomalous term which couples baryonic chiral partners to the
mesons $\eta$ and $\eta^{\prime}$. As a consequence, this study also shows
that $N(1535)$ is predominantly the chiral partner of $N(939)$, and
$\Lambda(1670)$ the chiral partner of $\Lambda(1116)$.

\section{Axial anomaly in the light mesonic sector}

Let us start with the (pseudo)scalar sector. For three flavors ($N_{f}=3$) the
corresponding mesonic matrix $\Phi$ (with elements $\Phi^{ij}=\overline
{q}_{\mathrm{R}}^{j}q_{\mathrm{L}}^{i}$) reads
\begin{equation}
\Phi=S+iP=\frac{1}{\sqrt{2}}\left(
\begin{array}
[c]{ccc}%
\frac{\sigma_{N}+a_{0}^{0}}{\sqrt{2}} & a_{0}^{+} & K_{0}^{\ast+}\\
a_{0}^{-} & \frac{\sigma_{N}-a_{0}^{0}}{\sqrt{2}} & K_{0}^{\ast0}\\
K_{0}^{\ast-} & K_{0}^{\ast0} & \sigma_{S}%
\end{array}
\right)  +\frac{1}{\sqrt{2}}\left(
\begin{array}
[c]{ccc}%
\frac{\eta_{N}+\pi^{0}}{\sqrt{2}} & \pi^{+} & K^{+}\\
\pi^{-} & \frac{\eta_{N}-\pi^{0}}{\sqrt{2}} & K^{0}\\
K^{-} & \bar{K}^{0} & \eta_{S}%
\end{array}
\right)  \text{ ,} \label{phinf3}%
\end{equation}
where $a_{0}\equiv a_{0}(1450)$ and $K_{0}^{\ast}$ $\equiv K_{0}^{\ast}%
(1430)$. In addition, $\sigma_{N}$ corresponds predominantly to $f_{0}(1370)$
and $\sigma_{S}$ to $f_{0}(1500)$ (with admixture among each other and with a
bare scalar glueball in $f_{0}(1710)$ \cite{dick}). In the pseudoscalar
sector, $\vec{\pi}$ are the pions and $K^{\pm},K^{0},\bar{K}^{0}$ the kaons,
while $\eta_{N}\equiv\sqrt{1/2}(\bar{u}u+\bar{d}d)$ and $\eta_{S}\equiv\bar
{s}s$ are the nonstrange and strange counterparts of $\eta\equiv\eta(547)$ and
$\eta^{\prime}\equiv\eta^{\prime}(958),$ see below.

By applying the chiral transformation $SU(3)_{R}\times SU(3)_{L}\times
U(1)_{A}$ (which reads $q_{\mathrm{L}/\mathrm{R}}\rightarrow(\mathrm{e}%
^{\mp\mathrm{i}\alpha/2}U_{\mathrm{L}/\mathrm{R}})q_{\mathrm{L}/\mathrm{R}}$,
$\alpha$ refers to $U(1)_{A}$), the matrix $\Phi$ transforms as
\begin{equation}
\Phi\rightarrow e^{-i\alpha}U_{L}\Phi U_{R}^{\dagger}\text{ ,}%
\end{equation}
hence the name \textit{heterochiral: }it picks up both matrices $U_{L}$ and
$U_{R}$. Generally, terms of a chiral model (e.g. \cite{dick,rosenzweig}) are
powers of $\Phi^{\dagger}\Phi$, such as $\mathrm{tr}(\Phi^{\dagger}\Phi)$,
$\mathrm{tr}(\Phi^{\dagger}\Phi)^{2}$, \textit{..., }thus invariant under the
chiral transformation $SU(3)_{R}\times SU(3)_{L}$ and also under the axial
transformation $U(1)_{\mathrm{A}}$. However, there are other possibilities:
one can use the property of the determinant that under $SU(3)_{R}\times
SU(3)_{L}\times U(1)_{A}$ transforms as
\begin{equation}
\mathrm{det}(\Phi)=\frac{1}{6}\varepsilon^{ijk}\varepsilon^{i^{\prime
}j^{\prime}k^{\prime}}\Phi^{ii^{\prime}}\Phi^{jj^{\prime}}\Phi^{kk^{\prime}%
}\,\rightarrow\mathrm{e}^{-3\mathrm{i}\alpha}\mathrm{det}(\Phi)\,
\end{equation}
in order to build anomalous terms invariant under $SU(3)_{R}\times SU(3)_{L}$
but not under $U(1)_{A}$, such as:
\begin{equation}
\mathcal{L}_{\Phi}^{\text{anomaly}}=-a_{\mathrm{A}}^{(3)}[\mathrm{det}%
(\Phi)-\mathrm{det}(\Phi^{\dagger})]^{2}+\ldots\, \label{l1}%
\end{equation}
(dots refer to other terms \cite{anomaly}). Here, for simplicity we keep only
the \textquotedblleft third\textquotedblright\ term of that work (by using the
same notation). This term gives a contribution to the effective potential of
the form $-\alpha_{\mathrm{A}}\eta_{0}^{2}=-\alpha_{\mathrm{A}}(\sqrt{2}%
\eta_{N}+\eta_{S})^{2}$ with $\alpha_{\mathrm{A}}\simeq a_{\mathrm{A}}%
^{(3)}\phi_{N}^{2}$ . This term, which was also obtained in\ Ref.
\cite{rosenzweig} and naturally arises when integrating out a pseudoscalar
glueball \cite{psg}, clearly affects the mixing in the pseudoscalar sector
(but not in the scalar sector). The physical fields $\eta\equiv\eta(547)$ and
$\eta^{\prime}\equiv\eta^{\prime}(958)$ are
\[
\eta=\eta_{N}\cos\theta_{P}~+\eta_{S}\sin\theta_{P}\ ,\eta^{\prime}=-\eta
_{N}\sin\theta_{P}~+\eta_{S}\cos\theta_{P}\text{ .}%
\]
The pseudoscalar mixing angle $\theta_{P}$ reads \cite{feldmann,anomaly}:
\begin{equation}
\theta_{P}=-\frac{1}{2}\arctan\left[  \frac{4\sqrt{2}\alpha_{\mathrm{A}}%
}{2(m_{K}^{2}-m_{\pi}^{2}-\alpha_{\mathrm{A}})}\right]  \text{ ;}%
\end{equation}
$\theta_{P}$ is negative for realistic values of $\alpha_{\mathrm{A}}$. The
term of Eq. (\ref{l1}) alone is not sufficient for a precise description of
the axial anomaly, but allows to understand its most salient phenomenological
features. Numerically, $\theta_{P}$ varies between $-40^{\circ}$
and$-45^{\circ}$~\cite{feldmann}.

Next, we move to (axial-)vector mesons, described by the matrices $V_{\mu}$
(with $J^{PC}=1^{--}$ , corresponding to the resonances \{$\rho(770)$,
$K^{\ast}(892)$, $\omega(782)$, $\phi(1020)\}$) and $A_{\mu}$ (with
$J^{PC}=1^{++},$ $\{a_{1}(1260)$, $K_{1,A}$, $f_{1}(1285)$, $f_{1}(1420)\}$),
see also the PDG for details \cite{pdg}.\ The chiral objects are the
right-handed and the left-handed currents $R_{\mu}^{ij}=\bar{q}_{\mathrm{R}%
}^{j}\gamma_{\mu}q_{\mathrm{R}}^{i}$, $L_{\mu}^{ij}\equiv\bar{q}_{\mathrm{L}%
}^{j}\gamma_{\mu}q_{\mathrm{L}}^{i}\,$ with $R_{\mu}=V_{\mu}-A_{\mu}\,$and
$L_{\mu}=V_{\mu}+A_{\mu}.$ Under $SU(3)_{R}\times SU(3)_{L}\times U(1)_{A}$:
\begin{equation}
L_{\mu}\longrightarrow U_{\mathrm{L}}\,R_{\mu}\,U_{\mathrm{L}}^{\dagger
}\,,\text{ }R_{\mu}\longrightarrow U_{\mathrm{R}}\,R_{\mu}\,U_{\mathrm{R}%
}^{\dagger}\,,
\end{equation}
therefore these multiplets are named \textit{homochiral} (either only
$U_{\mathrm{L}}$ or only $U_{\mathrm{R}}$ enter, respectively). Here it is not
possible to write down a term such as in\ Eq. (\ref{l1}). (Other more
complicated Wess-Zumino terms, see e.g. Ref. \cite{wz}, exist but to not
affect the isoscalar mixing). As a consequence, we expect the isoscalar mixing
to be much suppressed. This is in very good agreement with observations: the
resonance $\omega(782)$ is almost purely nonstrange, while $\phi(1020)$ almost
purely strange. Similarly, $f(1285)$ is predominantly nonstrange and
$f_{1}(1420)$ strange.

Ground-state \textit{tensor mesons (}$J^{PC}=2^{++},$ $\{a_{2}(1320)$,
$K_{2}^{\ast}(1430)$, $f_{2}(1270),$ $f_{2}^{\prime}(1525)\}$), represent also
a very well-known nonet of $\bar{q}q$ states \cite{tensor}. Together with
their not yet known chiral partners ($J^{PC}=2^{--},$ only $K_{2}(1820)$ has
been discovered), they are described by the heterochiral multiplet $L_{\mu\nu
}^{ij}=\bar{q}_{\mathrm{L}}^{j}(\gamma_{\mu}\overleftrightarrow{D_{\nu}%
}+\gamma_{\nu}\overleftrightarrow{D_{\mu}}+\ldots)q_{\mathrm{L}}^{i}$ ,
$R_{\mu\nu}^{ij}\equiv\bar{q}_{\mathrm{R}}^{j}(\gamma_{\mu}\overleftrightarrow
{D_{\nu}}+\gamma_{\nu}\overleftrightarrow{D_{\mu}}+\ldots)q_{\mathrm{R}}%
^{i}\,$with $D_{\mu}$ being the covariant derivative. (The tensor states are
$(L_{\mu\nu}+R_{\mu\nu})/2$). The currents transform as $L_{\mu\nu
}\longrightarrow U_{\mathrm{L}}\,L_{\mu\nu}\,U_{\mathrm{L}}^{\dagger}\,$and
$R_{\mu\nu}\longrightarrow U_{\mathrm{L}}\,R_{\mu\nu}\,U_{\mathrm{L}}%
^{\dagger}$, hence we are again in presence of a \textit{homochiral}
multiplet. Just as before, we expect a small isoscalar mixing, a fact which is
very well confirmed by experiments: $f_{2}(1270)$ is to a very good extent
purely nonstrange and $f_{2}^{\prime}(1525)$ strange.

We now move to \textit{heterochiral vectors}, which contain the pseudovector
states $P_{\mu}\equiv\{b_{1}(1235)$, $K_{1,B}$, $h_{1}(1170)$, $h_{1}(1380)\}$
with $J^{PC}=1^{+-}$and their chiral partners, the orbitally excited vector
mesons $S_{\mu}\equiv\{\rho(1700)$, $K^{\ast}(1680)$, $\omega(1650),\phi
(???)\}$ with $J^{PC}=1^{--}$ (see also Ref. \cite{piotrowska} where the
identification $\phi(???)\equiv\phi(1930)$ was put forward). To this end, we
construct the object $\Phi_{\mu}^{ij}\equiv\bar{q}_{\mathrm{R}}^{j}%
(\overleftrightarrow{D_{\mu}}+\ldots)q_{\mathrm{L}}^{i}\,=S_{\mu}%
+\mathrm{i}P_{\mu},$ which transforms as $\Phi_{\mu}\rightarrow\mathrm{e}%
^{-\mathrm{i}\alpha}\,U_{\mathrm{L}}\,\Phi_{\mu}\,U_{\mathrm{R}}^{\dagger}$,
i.e. just as (pseudo)scalar mesons. The anomalous Lagrangian (keeping only the
term analogous to Eq. (\ref{l1})) reads
\begin{equation}
\mathcal{L}_{\Phi_{\mu}}^{\text{anomaly}}=-b_{\mathrm{A}}^{(3)}(\varepsilon
^{ijk}\varepsilon^{i^{\prime}j^{\prime}k^{\prime}}\Phi^{ii^{\prime}}%
\Phi^{jj^{\prime}}\Phi_{\mu}^{kk^{\prime}}-h.c.)^{2}\,+...,
\end{equation}
which reduces to $-\beta_{\mathrm{A}}(\sqrt{2}h_{1,N}^{\mu}+h_{1,S}^{\mu}%
)^{2},$ with $\beta_{\mathsf{A}}\simeq b_{\mathrm{A}}^{(3)}\phi_{N}^{2},$ when
quadratic terms are considered. Here, a large mixing between the nonstrange
$h_{1,N}^{\mu}$ and the strange $h_{1,S}^{\mu}$ components is possible.
Unfortunately, the present experimental knowledge of the physical states is
very poor ($h_{1}(1380)$ is still omitted from the summary of the PDG), thus a
verification of their mixing is not possible. It will be interesting in the
future to study these states in more detail.

Last, and most interestingly, we study the \textit{heterochiral tensors. }They
include the nonet of pseudotensor mesons $P_{\mu\nu}\equiv\{\pi_{2}%
(1670),K_{2}(1770),\eta_{2}(1645),\eta_{2}(1870)\}$ with $J^{PC}=2^{-+}$ (see
also\ Ref. \cite{wang}) as well as their not yet known chiral partners
$S_{\mu\nu}$ with $J^{PC}=2^{++}$. The chiral object is $\Phi_{\mu\nu}%
=S_{\mu\nu}+\mathrm{i}P_{\mu\nu}\equiv\bar{q}_{\mathrm{R}}^{j}%
(\overleftrightarrow{D_{\mu}}\overleftrightarrow{D_{\nu}}+\ldots
)q_{\mathrm{L}}^{i},\,$which transforms as $\Phi_{\mu\nu}\rightarrow
\mathrm{e}^{-\mathrm{i}\alpha}\,U_{\mathrm{L}}\,\Phi_{\mu\nu}\,U_{\mathrm{R}%
}^{\dagger},$ thus just as the (pseudo)scalars. The anomalous Lagrangian
(keeping only the term analogous to Eq. (\ref{l1})) reads%

\begin{equation}
\mathcal{L}_{\Phi_{\mu\nu}}^{\text{anomaly}}=c_{\mathrm{A}}^{(3)}%
(\varepsilon^{ijk}\varepsilon^{i^{\prime}j^{\prime}k^{\prime}}\Phi
^{ii^{\prime}}\Phi^{jj^{\prime}}\Phi_{\mu\nu}^{kk^{\prime}}-h.c.)^{2}%
+...\text{,}%
\end{equation}
that reduces to $-\gamma_{A}(\sqrt{2}\eta_{2,N}^{\mu\nu}+\eta_{2,S}^{\mu\nu
})^{2}$ with $\gamma_{A}\simeq c_{\mathrm{A}}^{(3)}\phi_{N}^{2}$ for quadratic
terms. Interestingly, the phenomenological study of Ref. \cite{adrian} found
that the physical states are
\[%
\begin{pmatrix}
\eta_{2}(1645)\\
\eta_{2}(1870)
\end{pmatrix}
=%
\begin{pmatrix}
\cos\theta_{PT} & \sin\theta_{PT}\\
-\sin\theta_{PT} & \cos\theta_{PT}%
\end{pmatrix}%
\begin{pmatrix}
\eta_{2,N}=\frac{\bar{u}u+\bar{d}d}{\sqrt{2}}\\
\eta_{2,S}=\bar{s}s
\end{pmatrix}
\text{ with }\theta_{PT}\simeq-42^{\circ},\,
\]
which is a surprisingly large mixing. This result can be nicely explained by
the axial anomaly being active in this sector. Moreover, the corresponding
mixing angle turns out to be negative (for realistic values of $\gamma
_{\mathrm{A}}$) just as in the (pseudo)scalar sector:%

\begin{equation}
\theta_{PT}\simeq-\frac{1}{2}\arctan\left[  \frac{4\sqrt{2}\gamma_{\mathrm{A}%
}}{2(m_{K_{2}(1770)}^{2}-m_{\pi_{2}(1660)}^{2}-\gamma_{\mathrm{A}})}\right]
<0\text{ .}%
\end{equation}
Again, such an expression is only approximate since other anomalous terms are
neglected, but the main point is that, just as for $\eta$ and $\eta^{\prime}$,
the anomaly allows for a large mixing. Future experimental work at the ongoing
Jefferson lab can investigate these resonances \cite{gluex}.

\section{Axial anomaly in the light baryonic sector}

The object $(\det\Phi-\det\Phi^{\dagger})$, properly coupled to baryons,
affects some decay channels, most notably $N(1535)\rightarrow N\eta.$ To this
end, the mirror assignment for the baryons multiplets is needed \cite{detar}.
Here, we shall use the $N_{f}=3$ version of the mirror assignment presented
in\ Ref. \cite{olbrich}. The physical fields are%
\begin{align*}
B_{N} &  \equiv\{N(939),\Lambda(1116),\Sigma(1193),\Xi(1318)\}\ ,\text{ }%
B_{M}\equiv\{N(1440),\Lambda(1600),\Sigma(1660),\Xi(1690)\}\ ,\\
B_{M\ast} &  \equiv\{N(1535),\Lambda(1670),\Sigma(1620),\Xi(?)\}\ ,\text{
}B_{N\ast}\equiv\{N(1650),\Lambda(1800),\Sigma(1750),\Xi(?)\}\,,
\end{align*}
with the splitting into chiral baryonic multiplets
\[
B_{N}=\frac{N_{1}-N_{2}}{\sqrt{2}}\ ,\ B_{N\ast}=\frac{N_{1}+N_{2}}{\sqrt{2}%
}\ ,B_{M}=\frac{M_{1}-M_{2}}{\sqrt{2}}\ ,\ B_{M\ast}=\frac{M_{1}+M_{2}}%
{\sqrt{2}}\ ,
\]
which transform under chiral transformations $SU(3)_{R}\times SU(3)_{L}$ as
\begin{equation}
N_{1R(L)}\rightarrow U_{R(L)}N_{1R(L)}U_{R}^{\dagger}\ ,\text{ }%
N_{2R(L)}\rightarrow U_{R(L)}N_{2R(L)}U_{L}^{\dagger}\text{ , }M_{1R(L)}%
\rightarrow U_{L(R)}M_{1R(L)}U_{R}^{\dagger}\ ,\text{ }M_{2R(L)}\rightarrow
U_{L(R)}M_{2R(L)}U_{L}^{\dagger}\nonumber
\end{equation}
The chirally symmetric but axial anomalous Lagrangian reads
\cite{baryonanomaly}:
\begin{align}
\mathcal{L}_{A}^{N_{f}=3}= &  \lambda_{A1}(\det\Phi-\det\Phi^{\dagger
})\mathrm{Tr}(\bar{M}_{1R}N_{1L}-\bar{N}_{1L}M_{1R}-\bar{M}_{2L}N_{2R}+\bar
{N}_{2R}M_{2L})\nonumber\\
+ &  \lambda_{A2}(\det\Phi-\det\Phi^{\dagger})\mathrm{Tr}(\bar{M}_{1L}%
N_{1R}-\bar{N}_{1R}M_{1L}-\bar{M}_{2R}N_{2L}+\bar{N}_{2L}M_{2R})\ .
\end{align}
Considering that $\det\Phi-\det\Phi^{\dagger}\propto i(\sqrt{2}\eta_{N}%
+\eta_{S})+\dots\ ,$ the anomaly gives contributions in which $B_{N}$ and
$B_{M_{\ast}}$ (as well as $B_{M}$ and $B_{N_{\ast}}$) couple to $\eta_{0}$,
therefore an enhanced decay of the type $B_{M_{\ast}}\rightarrow B_{N}\eta$
follows. In particular, the anomalously large decay $N(1535)\rightarrow N\eta$
show that $N(1535)$ and $N(939)$ are chiral partners. In addition, also the
decay $\Lambda(1670)\rightarrow\Lambda\eta$ and the coupling of $N(1535)$ to
$N\eta^{\prime}$ can be correctly described, see Ref. \cite{baryonanomaly} for
further details.

In the end, we recall that the axial anomaly is also important to describe the
decays of the not-yet discovered pseudoscalar glueball. In fact, this state
couples through the axial anomaly to both mesons and baryons
\cite{baryonanomaly,psg}.

\section{Conclusions}

In this work we have revisited the role of the axial anomaly for what concerns
the phenomenology of the light mesonic and baryonic sectors. In the mesonic
sector one can -for the so-called \textquotedblleft
heterochiral\textquotedblright\ multiplets \cite{anomaly}- write down
anomalous terms similar to the one which are usually introduced for
(pseudo)scalar mesons. Hence, a large strange-nonstrange mixing in the
pseudotensor nonet and in the pseudovector nonet are possible (in the former
some experimental evidence already exists \cite{adrian}). On the contrary, the
ground-state vector and tensor mesons are \textquotedblleft
homochiral\textquotedblright, therefore the isoscalar states are (almost
purely) nonstrange and strange, respectively.

In the baryonic sector, the axial anomaly can help to understand some decay
processes involving the $\eta$ meson, most notably $N(1535)\rightarrow N\eta$
\cite{baryonanomaly} In fact, an anomalous term which connects baryonic chiral
partners to $\eta$ and $\eta^{\prime}$ is possible when the mirror assignment
is used. In turn, this approach shows that $N$ and \ $N(1535)$ are
predominantly chiral partners.

\bigskip

\textbf{Acknowledgments}: the author thanks R. Pisarski, A. Koenigstein, L.
Olbrich, M. Zetenyi, and D.\ H. Rischke for cooperation. Financial support
from the Polish National Science Centre NCN through the OPUS project
no.~2015/17/B/ST2/01625 and from the DFG under grant no.\ RI 1181/6-1 are aknnowledged.

\end{document}